\documentclass[aps,preprint,onecolumn,showpacs,amsmath,amssymb]{revtex4}
\usepackage{graphics}

\begin{document}

\title{Decoherence as a signature of an excited state quantum phase transition}

\author{A. Rela\~{n}o$^{1 }$}
\email{armando@iem.cfmac.csic.es}
\author{ J. M. Arias$^{2}$}
\author{ J. Dukelsky$^{1}$}
\author{ J. E. Garc\'{\i}a-Ramos$^{3}$}
\author{ P. P\'erez-Fern\'andez$^{2}$}

 \affiliation{$^{1}$ Instituto de Estructura de la Materia, CSIC, Serrano 123,
E-28006 Madrid, Spain \\
$^{2}$ Departamento de F\'{\i}sica At\'omica, Molecular y
  Nuclear, Facultad de F\'{\i}sica, Universidad de Sevilla,
  Apartado~1065, 41080 Sevilla, Spain \\
$^{3}$ Departamento de F\'{\i}sica Aplicada, Universidad de Huelva, 21071 Huelva, Spain}

\date{\today}
\begin{abstract}
We analyze the decoherence induced on a single qubit by the interaction with a two-level boson system with
critical internal dynamics. We explore how the decoherence process is affected by the presence of quantum phase
transitions in the environment. We conclude that the dynamics of the qubit changes dramatically when the
environment passes through a continuous excited state quantum phase transition. If the system-environment coupling energy
equals the energy at which the environment has a critical behavior, the decoherence induced on the qubit is
maximal and the fidelity tends to zero with finite size scaling obeying a power-law.
\end{abstract}


 \maketitle

Real quantum systems always interact with the environment. This interaction leads to decoherence, the process by
which quantum information is degraded and purely quantum properties of a system are lost. Decoherence provides a theoretical basis for the quantum-classical transition \cite{Zurek:03},
emerging as a possible explanation of the quantum origin of the classical world. It is also a major obstacle for building a quantum computer \cite{Nielsen} since it can produce the loss of the quantum
character of the computer. Therefore, a complete characterization of the decoherence process and its relation with
the physical properties of the system and the environment is needed for both fundamental and practical purposes.

The connection between decoherence and environmental quantum phase transitions 
has been recently investigated \cite{Zan:06, Paz:07, Paz:08}. A universal Gaussian decay regime in the fidelity of
the system was initially identified, and related to a second-order quantum phase transition in the environment
\cite{Paz:07}; as a consequence, the decoherence process was postulated as an indicator of a quantum phase
transition in the environment. Subsequently, this analysis was refined, and it was found that the universal regime
is neither always Gaussian \cite{Rossini:07}, nor  always related to an environmental quantum phase transition
\cite{Paz:08}.

In this paper we analyze the relationship between decoherence and an environmental excited state quantum phase
transition (ESQPT). We show that the fidelity of a single qubit, coupled to a two-level boson
environment, becomes singular when the system-environment coupling energy equals the critical energy for the
occurrence of a continuous ESQPT in the environment. Therefore, our results establish that
a critical phenomenon in the environment entails a singular behavior in the decoherence induced in the central
system.

\begin{figure}
\begin{center}
\rotatebox{-0}{\scalebox{0.3}[0.3]{\includegraphics{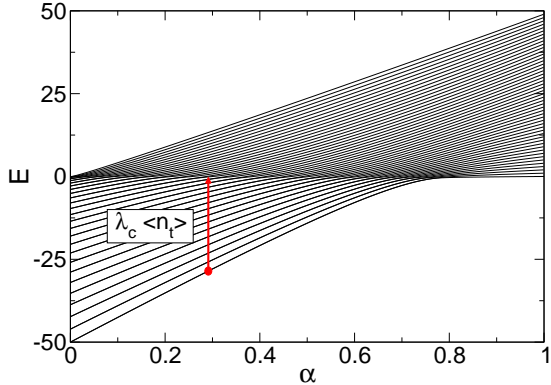}}} \caption[]{(Color online). Energy levels for the environment Hamiltonian (\ref{eq:twolevel}) with $N=50$. The arrow (red online) shows the jump that the coupling with the central qubit produces in the environment.} \label{fig:state}
\end{center}
\end{figure}

An ESQPT is a nonanalytic evolution of some excited states of a system as the Hamiltonian control parameter is
varied. It is analogous to a standard quantum phase transition (QPT), but taking place in some excited state of
the system, which defines the critical energy $E_c$ at which the transition takes place. We can distinguish between different kinds of ESQPT. As it is stated in \cite{Cejnar:07},
in the thermodynamic limit a crossing of two levels at $E=E_c$ determines a first order ESQPT, while if the number of interacting
levels is locally large at $E=E_c$ but without real crossings, the ESQPT is continuous. In this paper we will concentrate
in the latter case, which usually entails a singularity in the density of states (for an illustration see Fig.
\ref{fig:state}). As the entropy of a quantum system is related to its density of states, a relationship between
an ESQPT and a standard phase transition at a certain critical temperature, can be established  in the
thermodynamic limit \cite{Pavel}.

These kinds of phase transitions have been identified in the Lipkin model \cite{Heiss:05}, in the interacting
boson model \cite{Heinze:06}, and in more general boson or fermion two-level pairing Hamiltonians (for a complete
discussion, including a semiclassical analysis, see \cite{Caprio:08}). In all these cases, the ESQPT takes place
beyond the critical value of the Hamiltonian control parameter, implying that the critical point moves from the
ground state to an excited state.

Here we consider an environment having both QPTs
and ESQPTs coupled to a single qubit. The Hamiltonian of the environment, defined as a function of a control
parameter $\alpha$, presents a QPT at a critical value $\alpha_c$. We define a coupling between the central qubit and the
environment that entails an effective change in the control parameter, $\alpha \rightarrow
\alpha'$, making the environment to cross the critical point if $\alpha' >\alpha_c$. Moreover, the
coupling also implies an energy transfer to the enviroment $E \rightarrow E'$, and therefore it can also make the
environment to reach the critical energy $E_c$ of an ESQPT.

Following \cite{Paz:07} we will consider our system composed by a spin $1/2$ particle coupled to a spin
environment by the Hamiltonian $H_{SE}$:
\begin{equation}
H_{SE} = I_S \otimes H_E + \left|0\right> \left<0\right| \otimes H_{\lambda_0} + \left|1\right> \left<1\right| \otimes
H_{\lambda_1},
\end{equation}
where $\left|0\right>$ and $\left|1\right>$ are the two components of the spin $1/2$ system, and $\lambda_0$, $\lambda_1$ the couplings of each
component to the environment. The three terms $H_E$, $H_{\lambda_0}$ and $H_{\lambda_1}$ act on the Hilbert space of the
environment.

With this kind of coupling, the environment evolves with an effective Hamiltonian depending on the state of the
central spin $H_j = H_E + H_{\lambda_j}$, $j=0,1$. Assuming $\hbar=1$, it gives rise to a decoherence factor $r(t) = \left< \Psi(0) \right| e^{i H_0 t} e^{-i H_1 t} \left| \Psi(0) \right>$,
%
where $\Psi(0)$ represents a generic environmental state at $t=0$. If the environment is initially in its ground
state $\left| g_0 \right>$, the decoherence factor is determined, up to an irrelevant phase factor, by $H_1$, and its absolute value is equal to
\begin{equation}
\left|r(t)\right| = \left|\left< g_0 \right| e^{-i H_1 t} \left| g_0 \right>\right|.
\end{equation}
This quantity has the same form as the Loschmidt echo or the fidelity, and it contains all the relevant
information about the decoherence process.

To be specific, let us consider a spin environment described by the well known Lipkin model,
\begin{equation}
H_E=\alpha \left( \frac{N}{2} + \sum^N _{i=1} S^z_i \right)- \frac{4(1-\alpha)}{N} \sum^N_{i,j=1} S^x_i S^x_j ~,
\label{eq:Lipkin}
\end{equation}
where $S^x_i$, $S^y_i$, and  $S^z_i$, are the three components of a spin $1/2$ in the site $i$ of a spin chain
with $N$ sites, and $\alpha$ is a control parameter given in arbitrary units of energy. Consequently, in the following energy and time are always written in the corresponding arbirtary units.
%
%

Using the Schwinger representation of the spin operators we transform the Hamiltonian (\ref{eq:Lipkin}) into a two
level boson Hamiltonian, constructed out of scalar $s$ and $t$ bosons of opposite parity
\begin{equation}
H_E=\alpha n_t - \frac{1-\alpha}{N} (Q_t)^2,~~ {\rm with}~~ Q_t= s^{\dagger} t + t^{\dagger} s,
\label{eq:twolevel}
\end{equation}
where $n_t$ is the number of $t$ bosons and $N$ the total number of
bosons. This particular Hamiltonian belongs to a more general class of
two-level Hamiltonian that has been extensively applied to nuclear
structure \cite{Vidal:06,Bonatsos:08} and molecular physics
\cite{Curro:08}, and recently also proposed as a model for optical cavity QED \cite{Morrison:08}.

This Hamiltonian has a second order QPT at $\alpha_c=4/5$ \cite{Vidal:06}. The critical point can be easily calculated
in the thermodynamic limit assuming a condensed boson of the form $\Gamma^{\dagger}=\left( s^{\dagger}+\beta
t^{\dagger} \right) / \sqrt{1 + \beta^2}$. For $\alpha>4/5$ the environment is a condensate of $s$ bosons ($\beta=0$)
corresponding to a ferromagnetic state in the spin representation. For $\alpha<4/5$ the environment condensate mixes
$s$ and $t$ bosons breaking the reflection symmetry ($|\beta| >0$).

Choosing $\lambda_0=0$ and $\lambda_1=\lambda$ the coupling Hamiltonian reduces to a very simple form $H_{Coup}= \lambda n_t$,
%
%
which results into the effective Hamiltonians for each component of the systems
\begin{eqnarray}
H_0 &=& \alpha n_t - \frac{1-\alpha}{N} (Q_t)^2, \\
H_1 &=&(\alpha+\lambda) n_t - \frac{1-\alpha}{N} (Q_t)^2~.
\end{eqnarray}
Therefore, the system-environment coupling parameter $\lambda$ modifies the environment Hamiltonian. Using the coherent
state approach \cite{Vidal:06}, it is straightforward to show that $H_1$ goes through a second order QPT at $\lambda_*=4
- 5\alpha$, for $\alpha < 4/5$. Furthermore, a semiclassical calculation \cite{Caprio:08} shows that $H_E$ also passes through an ESQPT at
$E_c=0$, if $\lambda < \lambda_*$. This phenomenon is illustrated in Fig. \ref{fig:state}. For $\alpha<\alpha_c$, a bunch of energy
levels collapse at $E=0$, and thus a singularity in the density of states arises. At $\alpha=\alpha_c$, the collapse takes
place in the ground state, transforming the ESQPT into a standard QPT. For $\alpha>\alpha_c$ no singular behavior is
observed.

To understand how this critical energy can be reached when we couple this environment to a central qubit let us
take in consideration the following. We start the evolution with the ground state of the environment $\left| g_0
\right>$. At $t=0$ we switch on the interaction between the system and the environment, and let the system
evolve under the complete Hamiltonian. By instantaneously switching on this interaction, the energy of the
environment increases, and its state gets fragmented into a region with average energy equal to $E = \left< g_0
\right| H_1 (\alpha) \left| g_0 \right>$. Therefore, if $\left< g_0 \right| H_1 (\alpha) \left| g_0 \right> = 0$,
the coupling with the central qubit induces the environment to {\it jump} into a region arround the critical energy
$E_c$. This is illustrated in Fig. \ref{fig:state}. Starting from a state in the parity broken phase with $\alpha
< \alpha_c$, the coupling with the qubit, $H_{\lambda_1} = \lambda n_t$ increases the energy of the environment up
to the critical point $E_c$. Resorting to the coherent state approach \cite{Vidal:06}, we can obtain a critical
value of the coupling strength
\begin{equation}
\lambda_c (\alpha) = \frac{1}{2} \left( 4 - 5 \alpha \right), \; \; \; \alpha < \frac{4}{5}. \label{eq:zc}
\end{equation}

\begin{figure}
\begin{center}
\rotatebox{-90}{\scalebox{0.25}[0.25]{\includegraphics{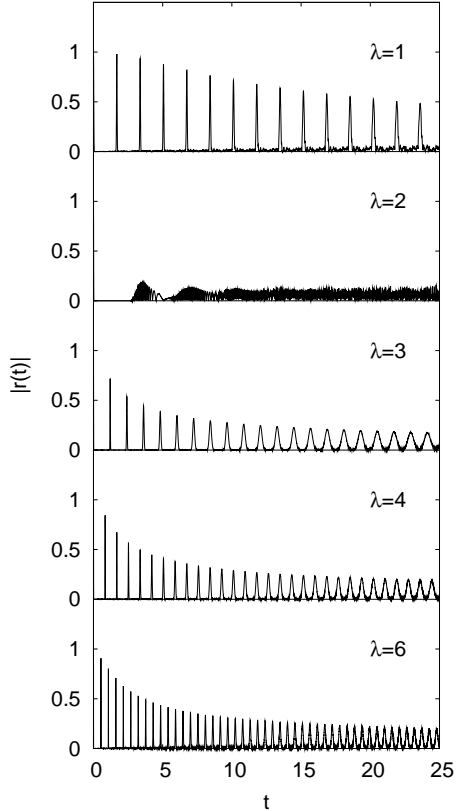}}} \caption[]{$|r(t)|$ for $\alpha=0$ and five
different values of $\lambda$. In all cases $N=10000$.} \label{fig:r(t)}
\end{center}
\end{figure}

In Fig. \ref{fig:r(t)} we show the modulus of the decoherence factor $\left| r(t) \right|$ for $\alpha=0$ and several
values of $\lambda$ (see caption). In four of the five cases we can see a similar pattern, fast oscillations plus a
smooth decaying envelope. For $\lambda \gtrsim \lambda_*$, this envelope is weakly dependent on $\lambda$. As we increase $\lambda$, the
frequency of the short period oscillations increases linearly $\nu \approx \lambda / 3$, but the main trend of the curve
remains the same. The shape of the envelope can be fitted to a Gaussian decay for short times, and to a power-law
decay for longer times; a similar behavior was identified in \cite{Paz:08} if the Hamiltonian parameter is close
to or larger than the critical value. However, the most striking feature of Fig. \ref{fig:r(t)} is the panel corresponding to
$\lambda=2$, for which $\left| r(t) \right|$ quickly decays to zero and then randomly oscillates around a small value.
We note that this particular case constitutes a singular point for both the shape of the envelope of $|r(t)|$ and
the period of its oscillating part. Making use of Eq. (\ref{eq:zc}) for $\alpha=0$ we obtain precisely $\lambda_c=2$, the
value at which the coherence of the system is completely lost. Therefore, the existence of an ESQPT in the
environment has a strong influence on the decoherence that it induces in the central system. We can summarize this
result with the following conjecture:

{\it If the system-environment coupling drives the environment to the critical energy $E_c$ of a continuous ESQPT, the
decoherence induced in the coupled qubit is maximal}.

\begin{figure}
\begin{center}
\rotatebox{-90}{\scalebox{0.3}[0.3]{\includegraphics{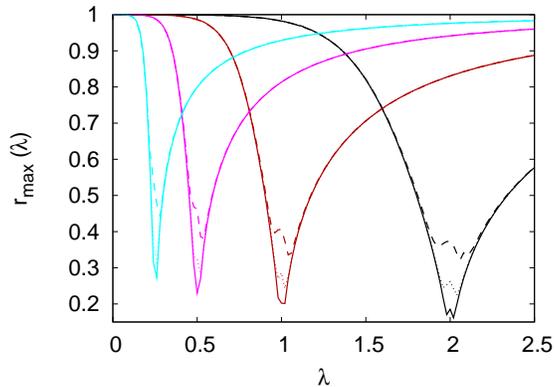}}} \caption[]{(Color online). $r_{max} (\lambda)$ in function of the
coupling $\lambda$, for different values of $\alpha$ and $N$. Black lines represent the case $\alpha=0$; dark grey (red online)
lines, $\alpha=0.4$; grey (magenta online), $\alpha=0.6$; and light grey (cyan online), $\alpha=0.7$. Solid lines represent
$N=10000$; dotted lines, $N=2500$; and dashed lines, $N=600$.} \label{fig:rmax_z}
\end{center}
\end{figure}

Let us check this conjecture for different values of $\alpha < \alpha_c=4/5$ . In Fig. \ref{fig:rmax_z} we show
how the decoherence process changes around the critical value $\lambda_c$ for different values
of $N$ and $\alpha$. As a representative quantity, we plot $r_{max} (\alpha)$ which is the value of $\left| r(t) \right|$ at the second maximum (the first
maximum is $\left| r(0) \right|=1$ in all cases). We can extract two main conclusions from Fig. \ref{fig:rmax_z}.
First, $r_{max}(\lambda)$ is minimum at $\lambda \approx \lambda_c$, as given in Eq. (\ref{eq:zc}) ---the four
cases plotted in the figure correspond to $\lambda_c= 0.25, \, 0.5, \, 1.0, \text{ and } 2.0$. Second, the
behavior of this quantity is smooth and independent of the environment size $N$, except in a small region around
$\lambda \approx \lambda_c$ where it becomes sharp and size dependent. Therefore, the decoherence
factor behaves in a critical way around $\lambda=\lambda_c$ where $r_{max} (\lambda_c)$ undergoes a dip towards
zero which is sharper and deeper for larger values of $N$.

\begin{figure}
\begin{center}
\rotatebox{-90}{\scalebox{0.3}[0.3]{\includegraphics{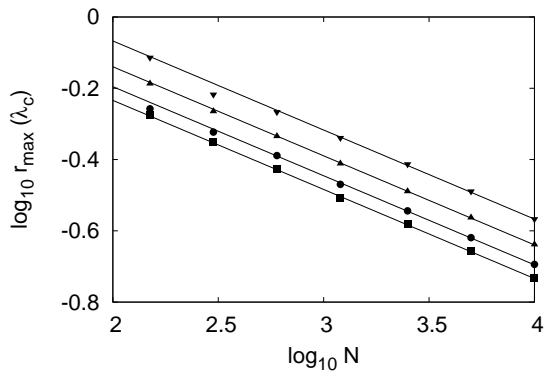}}} \caption[]{$r_{max} (\lambda_c)$ in function of the
size of the environment $N$, in a double logarithmic scale. Squares represent the case $\alpha=0$; circles, $\alpha=0.4$;
upper triangles, $\alpha=0.6$; lower triangles, $\alpha=0.7$. The straight lines correspond to a least squares fit to the
power law $r_{max} (\lambda_c) = A N^{-1/4}$.} \label{fig:rmax_N}
\end{center}
\end{figure}

On the light of these results, one may wonder how $r_{max} (\lambda_c)$ behaves in the thermodynamical limit. As it is
very difficult to do exact calculations beyond size $N > 10000$, we will rely on the finite size scaling analysis
to extrapolate its behavior for $N \rightarrow \infty$. In Fig. \ref{fig:rmax_N} we show how this quantity evolves with the size
$N$ of the environment. We plot results for the same values of $\alpha$ as in Fig. \ref{fig:rmax_z} in a double logarithmic scale
(see caption of Fig. \ref{fig:rmax_N} for details). In all the cases we obtain a power law $r_{max} (\lambda_c) \sim N^{-\gamma}$, with $\gamma=1/4$.
%
%

The results displayed in figures 3 and 4 confirm that the presence of an ESQPT in the environment spectrum is
clearly signaled by the qubit decoherence factor. Moreover, the quantity $r_{max}$ behaves in parallel way as the
order parameter $n_t$ at the ESQPT \cite{Caprio:08}.

Prior to the conclusions, it is worth mentioning that the environmental Hamiltonian Eq. (\ref{eq:twolevel}) is a
particular case of a more generic class of two-level bosonic systems \cite{Vidal:06}, described by $H=\alpha n_l -
(1-\alpha) \; Q^{(\omega)}_l \cdot Q^{(\omega)}_l / N$, with $Q^{(\omega)}_l = \left( s^{\dagger} \widetilde{l} +
l^{\dagger} s \right)^{(l)} + \omega \left( l^{\dagger} \widetilde{l} \right)^{(l)}$, where the boson
$s^{\dagger}$ is a scalar and the boson $l^{\dagger}$ has multipolarity $l$, as well as the operator
$Q^{(\omega)}_l$ depending on an extra parameter $\omega$. As a representative case, we have presented here
calculations using the environmental Hamiltonian with $l=0$ and $\omega=0$, characterized by a second order QPT.
For $l>1$ this wide class of Hamiltonians, that have been extensively used in nuclear and molecular physics,
present a richer structure with second order ($\omega=0$) and first order ($\omega \neq 0$) QPTs. A complete
analysis on the influence of the order of the ESQPT in the decoherence process will be given in a future
publication. Moreover, as the natural state of a realistic environment is a thermal one
---that is, a mixed state in which the environment is in contact with a thermal bath characterized by a temperature $T$--- it
will be also interesting to check to which extent the temperature affects this critical behavior.

In this paper, we have studied the decoherence induced on a one-qubit system by the interaction with a two-level
boson environment, which presents both a standard QPT and an ESQPT. Our main finding is that the decoherence is
maximal when the system-environment coupling introduces in the environment the energy required to undergo a continuous ESQPT.
The decoherence factor displays a critical behavior well described by $r_{max} (\lambda)$, the value of the second local
maximum of $|r(t)|$. $r_{max} (\lambda)$ approaches zero for $\lambda \rightarrow \lambda_c$ in the thermodynamical limit showing a
finite size scaling $r_{max} (\lambda_c) \propto N^{-\gamma}$, with a critical exponent $\gamma=1/4$. Therefore, we
conclude that the decoherence induced in a one-qubit system gives a unique signature of an ESQPT in the
environment. Finally, let us point out that quantum decoherence could constitute an important tool to detect
critical regions in the energy spectrum of mesoscopic systems; conversely, this knowledge could help to avoid
certain environments that are particularly efficient to destroy quantum coherence. Moreover, as the particular Hamiltonian we have studied in this paper has been broadly applied in nuclear structure, molecular physics and, more recently, optical cavity QED, it is feasible to use any of these systems as the environment in a decoherence experiment.

This work has been partially supported
by the Spanish Ministerio de Educaci\'on y Ciencia and by the European
regional development fund (FEDER) under projects number FIS2005-01105,  FIS2006-12783-C03-01
FPA2006-13807-C02-02 and FPA2007-63074, by Comunidad de Madrid and CSIC under
project 200650M012, and by Junta de
Analuc\'{\i}a under projects FQM160, FQM318, P05-FQM437 and
P07-FQM-02962. A.R. is supported by the Spanish program "Juan de la
Cierva", and P. P-F., by a grant from the Plan Propio of
the University of Sevilla.

\end{document}